\documentstyle [psfig,aps]{revtex}

\begin{document}
\draft
\title{Phenomenology of the Gowdy Universe on $T^3 \times R$  \thanks{
e-mail: berger@oakland.edu, garfinkl@oakland.edu}}
\author{Beverly K. Berger, David Garfinkle}
\address{Department of Physics, Oakland University, Rochester, MI 48309 USA}

\maketitle
\bigskip
\begin{abstract}
Numerical studies of the plane symmetric, vacuum
Gowdy universe on $T^3 \times R$ yield strong support for the conjectured
asymptotically
velocity term dominated (AVTD) behavior of its evolution toward the singularity
except, perhaps, at isolated spatial points. A generic solution is
characterized by
spiky features and apparent ``discontinuities'' in the wave amplitudes. It
is shown that
the nonlinear terms in the wave equations drive the system generically to
the ``small
velocity'' AVTD regime and that the spiky features are caused by the
absence of these
terms at isolated spatial points.
\end{abstract}
%\narrowtext
\section{Introduction}
Spatially inhomogeneous cosmologies constitute almost a ``terra incognita'' for
general relativity. In particular, the nature of the singularity in generic
cosmologies is unknown although Belinskii, Khalatnikov, and Lifshitz (BKL) have
conjectured that it is locally Mixmaster-like \cite{BKL}. Spatially homogeneous
universes' approach to the singularity can be either asymptotically
velocity term
dominated (AVTD) \cite{els,IM} or Mixmaster-like \cite{BKL,misner}. In
general, AVTD
singularities occur when the influence of spatial derivatives can be
neglected. This
means that the solution to the full Einstein equations eventually comes
arbitrarily
close to the solution to the truncated equations one obtains by eliminating
the terms
containing spatial derivatives
\cite{IM}. In spatially homogeneous models, spatial derivatives in the
metric create the
spatial scalar curvature which appears as a potential in minisuperspace
\cite{misner,ryan}. An AVTD homogeneous cosmology eventually evolves toward the
singularity as a Kasner solution. (The Kasner universe is an exact solution
of the
vacuum Einstein equations characterized by different expansion rates in
different
directions with flat spacelike hypersurfaces
\cite{kasner}.) Mixmaster dynamics describes the evolution (of non-flat
space-like
hypersurfaces) toward the singularity as an infinite sequence of
(approximately) Kasner
epochs. The Kasner indices (which determine the anisotropic expansion
rates) change
whenever the influence of the spatial scalar curvature reappears. Its recurring
influence means that the Mixmaster singularity is not AVTD. BKL claimed to prove
that the generic singularity in Einstein's equations and, in particular, in
spatially
inhomogeneous cosmologies is Mixmaster-like. While this result remains
controversial
\cite{bt}, it provides a conjecture which can be tested by evolving spatially
inhomogeneous collapsing universes numerically.

As part of such a
study of this issue \cite{bkbvm,bkb97c,bkb97d}, we have considered the
Gowdy universes on
$T^3 \times R$ as a test case. These models are one of the classes
discovered by Gowdy \cite{gowdy,bkb74,vm81} as a reinterpretation of
Einstein-Rosen waves
\cite{er}. While their plane symmetry precludes local Mixmaster behavior,
their nonlinear
gravitational wave interactions lead to an interesting phenomenology which
forms the subject of this paper. The Gowdy singularity has been conjectured
to be
asymptotically velocity term dominated (AVTD) \cite{gm1} and has been shown
rigorously to be so for the polarized case \cite{IM}. In addition, it has been
shown that for the polarized case, cosmic censorship holds: There exists a
natural foliation labeled by $0 \le \tau < \infty$ such that regular initial
data evolve without singularity for all $\tau < \infty$ where (in the collapse
direction) a curvature singularity occurs at $\tau = \infty$ \cite{cim}. Here we
shall use numerical studies to provide support for the extension of the
polarized Gowdy results to generic Gowdy models.

Gowdy cosmologies represent the simplest spatially inhomogeneous cosmological
solutions to Einstein's equations. The vacuum case may be interpreted as the
two polarizations of gravitational waves propagating in an inhomogeneous
background spacetime \cite{bkbvm,bkb74}. The model is plane symmetric with the
propagation direction of the waves orthogonal to the symmetry plane. Near the
singularity, the metric may be rewritten as locally Kasner with spatially
dependent
Kasner indices \cite{bkb74}. (This is a non-rigorous statement of AVTD
behavior.).
Matter fields compatible with the Gowdy metric have been discussed by
Carmeli, Charach,
and Malin \cite{cm,ccm}, and Ryan \cite{ryan-gowdy}. Here we shall consider
only vacuum
solutions.

The main phenomena of interest in the Gowdy universe are the growth of small
scale spatial structure and the approach to the AVTD regime. We shall
demonstrate that both result from the nonlinear terms in the wave equation for
$P$, the $+$ polarization of the gravitational waves. These act as space
and time
dependent potentials whose effect is to drive the system to velocity dominance.
As the AVTD regime is reached, the potentials approach zero. Their spatial
dependence means that wave amplitudes grow at some spatial points and
decrease at
others, eventually producing small scale spatial dependence in the
waveforms. Non-generic
behavior occurs at points where the coefficients of the potentials
(separately) vanish
identically. We note that this analysis in terms of potentials explains the
observed
episodic evolution of the system to the AVTD regime and the fact that the
conditions for
AVTD are reached at some spatial points before others. This should be
considered
together with the asymptotic expansion developed by Grubi\u{s}i\'{c} and Moncrief (GM)
\cite{gm1} which displays an exponential decay to the AVTD limit,
applicable in the last
episode.

The primary advantage of the Gowdy models as a test case for the study of
inhomogeneous cosmologies is the existence of variables in which the dynamical
equations for the wave amplitudes decouple from the constraints. Initial data
for the wave equations may be freely specified (restricted only in that the
total momentum along the direction of propagation must vanish) while the
constraints appear as prescriptions for the construction of the background
metric from the known solution to the dynamical equations. Thus the two
principal difficulties of numerical relativity---solving the initial value
problem and preserving the constraints during the evolution---become trivial.
This decoupling between dynamical and constraint equations disappears in more
complicated inhomogeneous models.  However, studies of $U(1)$ symmetric
cosmologies (with
only one Killing field) indicate that many features of the
Gowdy phenomenology such as spiky small scale spatial structure persist in
these more
complicated models \cite{bkbvm,bkbvm2,bkb97d} and, presumably, would be
important in the
dynamics of generic cosmologies.

In Section 2, the Gowdy model on $T^3 \times R$ will be reviewed. Section 3 will
contain a brief description of the numerical method. In Section 4, the
phenomenology will
be described. Section 5 will contain a unified explanation for the
phenomenology.
Discussion will be given in Section 6.

\section{The Model}
The Gowdy model
on $T^3 \times R$ is described by the metric \cite{gowdy,bkbvm}
\begin{eqnarray}
\label{gowdymetric}
ds^2&=&e^{{{-\lambda } \mathord{\left/ {\vphantom {{-\lambda
} 2}} \right. \kern-\nulldelimiterspace} 2}}e^{{\tau  \mathord{\left/
{\vphantom {\tau  2}} \right. \kern-\nulldelimiterspace} 2}}(-\,e^{-2\tau
}\,d\tau ^2+d\theta ^2)\nonumber \\
 &  &+e^{-\tau }\,[e^Pd\sigma ^2+2e^PQ\,d\sigma \,d\delta +(e^PQ^2+e^{-
P})\,d\delta ^2]
\end{eqnarray}
where $\lambda$, $P$, $Q$ are functions of $\theta$, $\tau$. The sign of
$\lambda$ differs from that in \cite{bkbvm} as required for consistency
with the form
given there for the equations for $\lambda$. The arguments in \cite{bkbvm} did not
depend on this error. We impose
$T^3$ spatial topology by requiring $0 \le \theta, \sigma, \delta \le 2 \pi$
and the metric functions to be periodic in $\theta$.
If we assume $P$ and $Q$ to be small, we find them to be respectively the
amplitudes of the $+$ and $\times$ polarizations of the gravitational
waves  with $\lambda$ describing the background in which they
propagate.  The time variable $\tau$ measures the area in the symmetry
plane with $\tau = \infty$ a curvature singularity. Einstein's equations split
into two groups. The first is nonlinearly coupled wave equations for $P$
and $Q$ (where $,_a = \partial / {\partial a}$):
\begin{equation}
\label{gowdywaveP}
P,_{\tau \tau }-\;e^{-\kern 1pt2 \tau }P,_{\theta \theta }-e^{2P}\left(
{Q,_\tau ^2-\;e^{-\kern 1pt2\tau }Q,_\theta ^2} \right)=0,
\end{equation}
\begin{equation}
\label{gowdywaveQ}
  Q,_{\tau \tau }-\;e^{-\kern 1pt2\tau }Q,_{\theta \theta }+\,2\,\left(
{P,_\tau Q,_\tau ^{}-\;e^{-\kern 1pt2\tau }P,_\theta Q,_\theta ^{}}
\right)=0.
\end{equation}
The second contains the Hamiltonian and $\theta$-momentum constraints
which can be expressed as first order equations for $\lambda$ in terms of
$P$ and $Q$:
\begin{equation}
\label{gowdyh0}
\lambda ,_\tau -\;[P,_\tau ^2+\;e^{-2\tau }P,_\theta ^2+\;e^{2P}(Q,_\tau
^2+\;e^{-2\tau }\,Q,_\theta ^2)]=0,
\end{equation}
\begin{equation}
\label{gowdyhq}
\lambda ,_\theta -\;2(P,_\theta P,_\tau +\;e^{2P}Q,_\theta Q,_\tau )=0.
\end{equation}
This split into dynamical and constraint equations removes two of the
most problematical areas of numerical relativity from this model:  (1) The
normally difficult initial value problem becomes trivial since $P$, $Q$
and their first time derivatives may be specified arbitrarily. The only
restriction, that the total $\theta$ momentum in the waves vanishes, is a
consequence of the requirement that metric variables be periodic in $\theta$.
(2) The constraints,  while guaranteed to be preserved in an analytic evolution
by the Bianchi identities, are not automatically preserved in a numerical
evolution with  Einstein's equations in differenced form. However, in
the Gowdy model,  the constraints are trivial since
$\lambda$ may be constructed from the  numerically determined
$P$ and $Q$. For the special case of the polarized Gowdy model ($Q=0$), $P$
satisfies a linear wave equation whose exact solution is well-known
\cite{bkb74}.  For this case, it has been proven that the approach to the
singularity is AVTD  
\cite{IM}.  This has also been conjectured to be true for generic Gowdy models
\cite{gm1}. We shall show in Section 5 how a numerical study of this model
provides strong support for this conjecture.

The wave
equations (\ref{gowdywaveP}) and (\ref{gowdywaveQ}) can be obtained by
variation of the
Hamiltonian \cite{vm81}
\begin{eqnarray}
\label{gowdywaveh}
H&=&{1 \over 2}\int\limits_0^{2\pi } {d\theta
\,\left[ {\pi _P^2+\kern 1pt\,e^{-2P}\pi _Q^2} \right]}\nonumber \\
  &+&{1 \over 2}\int\limits_0^{2\pi } {d\theta \,\left[ {e^{-
2\tau }\left( {P,_\theta ^2+\;e^{2P}Q,_\theta ^2} \right)}
\right]}=H_K+H_V .
\end{eqnarray}
The variation of the kinetic energy term, $H_K$, yields equations of motion for
the AVTD solution which arises when spatial derivatives are neglected. These
equations are exactly solvable with the solution
\begin{eqnarray}
\label{avtdeq}
P&=& {P_0} \; + \; \ln \left [ \cosh v \tau \, + \, \cos \psi \, \sinh v
\tau \right ] \;
\; \to v\tau \quad {\rm as}\;\tau \to \infty , \nonumber \\
Q&=&Q_0 \; + \; {{{e^{- {P_0}}} \, \sin \psi \, \tanh v \tau } \over {1 \,
+ \, \cos \psi
\, \tanh v \tau }}
\quad \quad \quad \;\to Q_\infty \quad {\rm as}\;\tau \to \infty , \nonumber
\\
\pi _P&=&v \; {{\tanh v \tau \, + \, \cos \psi } \over {1 \, + \, \cos \psi
\, \tanh
v \tau }}\quad \quad \quad \quad \quad
\quad \to v\quad \;\;{\rm as}\;\tau \to
\infty, \nonumber \\
\pi _Q&=& {e^{P_0}} \, v \, \sin \psi \equiv \pi_Q^0
\end{eqnarray}
given in terms of four functions of $\theta$: $\psi $, $v \ge 0$, $P_0$,
and $Q_0$.
Here ${Q_\infty} \equiv Q_0 + {e^{ - {P_0}}} \sin \psi / (1 +\cos \psi )$.
The large $\tau $ limits given above are for $\cos \psi \ne - \, 1$.  The
special case
$\cos \psi = - \, 1$ will be discussed later.
While (\ref{avtdeq}) may be solved for all these functions, it is
convenient to note that
\cite{gm1}
\begin{equation}
\label{vdef}
v = \sqrt{\pi_P^2 + \pi_Q^2 \, e^{-2P}}
\end{equation}
is the geodesic velocity in the target space (with metric $ds^2 = dP^2 +
e^{2P} dQ^2$)
of the wavemap \cite{vm81,gm1,bkbvm}.
If, as the singularity is approached, the behavior is AVTD, the
true solution will approach the AVTD one.  The exponential prefactor $e^{-2
\tau}$  in $H_V$ of (\ref{gowdywaveh}) makes plausible the conjectured AVTD
singularity. However, $P
\to v \tau$ (for $v > 0$) as $\tau \to \infty$. If $v > 1$, the term $V_2 =
e^{- 2 \tau} e^{2 P} Q,_{\theta}^2$ in $H_V$ can grow rather than
decay as $\tau \to \infty$. In fact, if one assumes the AVTD behavior as
$\tau \to
\infty$ in (\ref{avtdeq}), the wave equations become \cite{gm1}
\begin{eqnarray}
\label{PeqGM}
P,_{\tau \tau} &=& -e^{-2[1-v]\tau}[Q'_\infty ]^2+e^{-2\tau}v'' \tau
+\pi_Q^2 e^{-2 v \tau}\nonumber\\
\pi_Q,_\tau &=&e^{-2[1-v]\tau}[Q_\infty '' + 2 v' Q_\infty ' \tau]
\end{eqnarray}
where $' = d/d\theta$.
These equations are inconsistent if $v > 1$ unless $Q'_\infty = 0 =
Q''_\infty$,
which does not occur in a generic (unpolarized) Gowdy model. However, in
generic Gowdy
models, one also expects isolated points (a set of measure zero) with
$Q'_\infty = 0$ but
$Q''_\infty \ne 0$. In Section 5, we shall see that $v > 1$ can
persist at such
isolated values of $\theta$ where the AVTD conditions may not be satisfied.
GM thus conjecture that the AVTD limit as $\tau \to
\infty$ requires
$0 \le v < 1$ everywhere  except, perhaps, at a set of measure zero
(isolated values of
$\theta$) \cite{gm1}. We note that the polarized Gowdy model ($Q = \pi_Q =
0$) has the
AVTD solution $P = P_0 + \alpha \tau$, $\pi_P = \alpha$ with $\alpha(\theta)$
unrestricted in sign or magnitude \cite{bkb74,IM}. The fact that this
behavior is not a
limit of the generic AVTD solution (\ref{avtdeq}) illustrates further the
qualitative
distinction between the polarized case and $Q \ne 0$ but arbitrarily small.

\section{Numerical Methods}
The numerical method is discussed in detail elsewhere \cite{bkbvm}.
However, we shall
summarize it here. The work reported here was performed by using a
symplectic PDE
solver\cite{fleck,vm83}. Consider a system with one
degree of freedom described by $q(t)$ and its canonically conjugate momentum
$p(t)$ with a Hamiltonian
\begin{equation}
\label{1dofH}
H = {{p^2} \over {2m}} + V(q) = H_K + H_V.
\end{equation}
Note that the subhamiltonians $H_K$ and $H_V$ separately yield equations of
motion which are exactly solvable no matter the form of $V$. Variation of $H_K$
yields $\dot q = p/m$, $\dot p = 0$ with solution
\begin{equation}
\label{HKsoln}
p(t + \Delta t) = p(t) \quad, \quad q(t + \Delta t) = q(t) + {{p(t)} \over m}
\Delta t.
\end{equation}
Variation of $H_V$ yields $\dot q = 0$, $\dot p = -dV/dq$ with solution
\begin{equation}
\label{HVsoln}
q(t + \Delta t) = q(t) \quad, \quad p(t + \Delta t) = p(t) - \left. {{{dV}
\over {dq}}} \right|_t \, \Delta t.
\end{equation}
Note that the absence of momenta in $H_V$ makes (\ref{HVsoln}) exact for any
$V(q)$. One can then demonstrate that to evolve from $t$ to $t + \Delta t$ an
evolution operator ${\cal U}_{(2)}(\Delta t)$ can be constructed from the
evolution sub-operators ${\cal U}_K(\Delta t)$ and ${\cal U}_V(\Delta t)$
obtained from (\ref{HKsoln}) and (\ref{HVsoln}). One can show that
\cite{fleck}
\begin{equation}
\label{Uprescription}
{\cal U}_{(2)}(\Delta t) = {\cal U}_K(\Delta t/2)\,{\cal U}_V(\Delta
t)\,{\cal U}_K(\Delta t/2)
\end{equation}
reproduces the true evolution operator through order $(\Delta t)^2$. Suzuki has
developed a prescription to represent the full evolution operator to arbitrary
order \cite{suzuki}. For example
\begin{equation}
\label{4thorderU}
{\cal U}_{(4)}(\Delta t) ={\cal U}_{(2)}(s\Delta t)\,{\cal U}_{(2)}[(1-2s)\Delta
t]\,{\cal U}_{(2)}(s\Delta t)
\end{equation}
where ${s} = 1/(2 - {2^{1/3}})$. The advantage of Suzuki's approach is
that one only needs to construct ${\cal U}_{(2)}$ explicitly. ${\cal U}_{(2n)}$
is then constructed from appropriate combinations of ${\cal U}_{(2n-2)}$.

The generalization of this method to $N$ degrees of freedom and to fields is
straightforward. In the latter case, $V[\vec q(t)] \to V[\vec
q(\vec x,t)]$ so that $dV/dq$ becomes the functional derivative $\delta V /
\delta q$. On the computational spatial lattice, the derivatives that are
obtained in the expression for the functional derivative must be represented in
differenced form. We note that, to preserve $n$th order accuracy in time,
$n$th order accurate spatial differencing is required. Some discussion of this
has been given elsewhere \cite{bkbvm}.

Convergence studies have been performed to test the algorithm. We have
already shown
\cite{bkbvm} that at any fixed value of $\tau$, there exists a spatial
resolution at
which every spiky feature is resolved. Any finer resolution will yield identical
results. However, as we shall see, generic spiky features grow narrower as the
singularity is approached. This means (see \cite{bkbvm}) that a spatial
resolution which
was adequate at some $\tau_0$ will fail to resolve all features at some
$\tau_1 >
\tau_0$. This type of ``resolution dependence'' is well understood to be a
consequence
of the physical behavior of the model. If the physical system has features
which become
arbitrarily narrow as $\tau \to \infty$, no resolution will be adequate
everywhere for
all time.

\section{Gowdy Phenomenology}
Here we report the results of a numerical study of the evolution of generic
Gowdy models toward the singularity ($\tau \to \infty$). Rather than perform
simulations for hundreds of Gowdy models, we consider a single class of
models and argue that the observed behavior is generic. For the most part, we
shall use the initial data
$P = 0$, $\pi_P = v_0
\cos \theta$, $Q = \cos \theta$, and $\pi_Q = 0$.  This model is actually
generic for the following reasons:  The $\cos \theta$ dependence is the
smoothest nontrivial possibility.  Since we shall discuss the growth of
small scale spatial structure, we expect the cleanest effects to arise from
the smoothest initial data. A more complicated initial state will not yield
any qualitatively new phenomena. For example, choosing
$\cos n\theta$ instead yields the same solution repeated $n$ times on the
grid thus yielding the same result with  poorer resolution.  In addition, the
amplitude of
$Q$  is irrelevant since the Hamiltonian (\ref{gowdywaveh}) is invariant under
$Q \to \rho Q$, $P \to P - \ln \rho$ for any constant $\rho$.  This also means
that any unpolarized  model is qualitatively different from a polarized ($Q =
0$) one no matter how small $Q$. Traveling waves (subject to $\int {d\theta
\;\wp _\theta =0}$ where $\wp_\theta$ is the total $\theta$-momentum) yield
qualitatively similar behavior \cite{csthesis}. We are interested in the
spatial structure that can develop from smooth initial data. For this reason,
we have used $\cos \theta$ spatial dependence. Starting with a more complicated
wave form will yield an eventually more complicated spatial structure but
nothing qualitatively different will appear. Similarly, since rescaling $Q$
does not change the solution qualitatively, the amplitude of $Q$ is kept fixed.
In this discussion, we shall consider only $P$ and $Q$ and construct
$\lambda$ later if desired.

From the chosen class of initial data, with high spatial resolution, one
obtains after some time ($\tau \approx 12$) the profiles for $P$ and $Q$
shown in
Fig.~\ref{gowdyPQ}. The numbered features are typical for any generic
solution. The peaks labeled
\#1, \#2, and \#3 arise from the same features of the nonlinear
interactions. On a finer
spatial scale in Fig.~\ref{pqfine}, we notice that (a) the peak in $P$ is
associated with
an extremum of $Q$ (which may be a maximum or minimum) and (b) $|\pi_Q|$
becomes large with $\pi_Q < (> 0)$ for a maximum (minimum) in $Q$. The apparent
discontinuity in $Q$ (labeled \#4 in Fig.~\ref{gowdyPQ}) is shown on a finer
scale in
Fig.~\ref{qdisc}. We see that this feature occurs as $\pi_Q$ goes through
zero where $P <
0$. In Fig.~\ref{pqfine18}, the same features \#1 and \#4 are shown at
$\tau \approx 18$.
We see that the ``spikes'' have narrowed in $\theta$.

\begin{figure}[bth]
%fig5
\begin{center}
%\setlength{\unitlength}{1cm}
%\makebox[11.7cm]{\psfig{file=gpfig5.eps,width=9cm}}
\makebox[4in]{\psfig{file=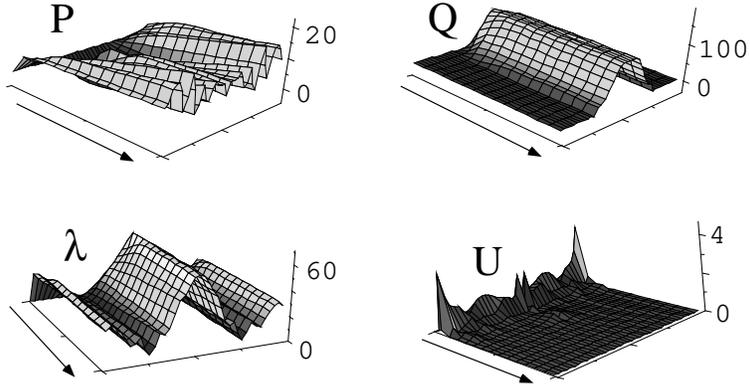,width=4in}}
{\protect \caption{$P$, $Q$, $\lambda$, and $U = e^{-
2\tau }\left( {P,_\theta ^2+\;e^{2P}Q,_\theta ^2} \right)$ are shown in the
$\theta$-$\tau$ plane. The arrow is in the direction of increasing $\tau$.}
\label{pqevol}}
\end{center}
\end{figure}

\begin{figure}[bth]
%fig6
\begin{center}
%\setlength{\unitlength}{1cm}
%\makebox[11.7cm]{\psfig{file=gpfig6.eps,width=9cm}}
\makebox[4in]{\psfig{file=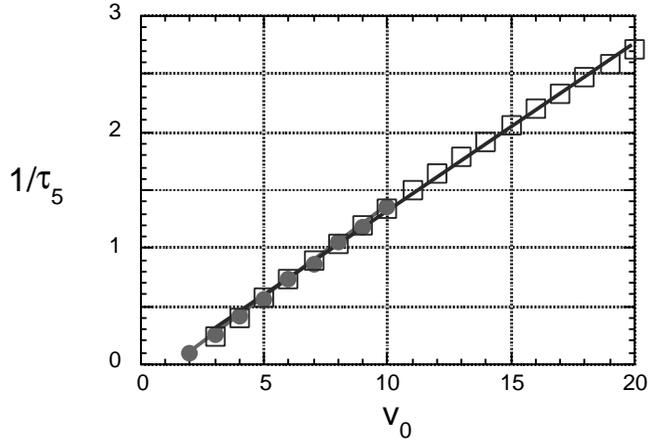,width=3.5in}}
\caption{Scaling in the Gowdy Model.  Plot of $1/\tau_5$, the 
inverse of the time at which the 5th peak appears
in $P$ vs $v_0$.  
Two cases are shown for initial data $P=0$, $Q= \cos \theta$:  
(1) $\pi_P = {1/ {\protect \sqrt{2}}}\  v_0 \cos \theta$, $\pi_Q ={1/
{\protect \sqrt{2}}}\  v_0
\cos
\theta$ is indicated by filled circles; (2) $\pi_P = v_0 \cos \theta$, $\pi_Q
= 0$ is indicated by open squares. The solid line is a linear best fit.}
\label{scaling}
\end{center}
\end{figure}

The evolution of $P$ and $Q$ is shown in Fig.~\ref{pqevol} (see also
\cite{bkbvm}) where
it is clearly seen that ever shorter wavelength modes develop until the point at
which the AVTD behavior sets in. Fig.~\ref{scaling} shows an interesting
self-similarity
in the early evolution. The parameter $v_0$ in the initial data is varied in a
sequence of simulations. In each case, the number of peaks in $P$ is counted
after each time step. We find
\begin{equation}
\label{selfsim}
{1 \over {\tau _N}}=a_N\left( {v_0-\bar v_0,_N} \right)
\end{equation}
where $\tau_N$ is the time at which $N$ peaks are first present in the waveform
and $a_N$ is a constant. From (\ref{selfsim}), $\bar v_0$ denotes the value
of $v_0$ such that the $N$th peak appears only at $\tau = \infty$. The scaling
shown in Fig.~\ref{scaling} is for $N = 5$. Similar (but less accurate)
straight line
fits are found for $N$ equal 3 and 7. (The wave equations are even in
$\theta$ as
are the initial data so that $P$ and $Q$ remain even functions of $\theta$.
Thus, except for a central peak, other peaks will form two at a time.)

Thus, one may summarize that spatial structure forms more rapidly and evolves
a more complicated form, the greater the initial ``energy'' in the waves as
denoted by, e.g., $v^2$ from (\ref{vdef}). Eventually, the waveforms of $P$
and $Q$
``freeze'' to the AVTD behavior of $P$ increasing linearly and $Q$ constant
in $\tau$.
As has been emphasized by Hern and Stuart (HS) \cite{hs}, the duration in
$\tau$ of the
observed non-AVTD regime depends on the spatial resolution of the simulation.

\begin{figure}[bth]
%fig7
\begin{center}
\makebox[4in]{\psfig{file=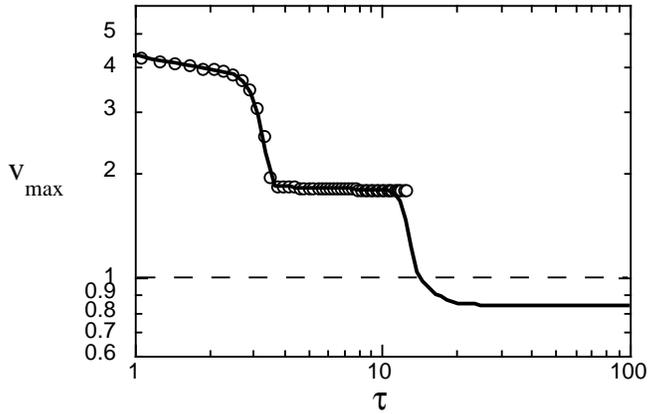,width=3.5in}}
%\makebox[11.7cm]{\psfig{file=gpfig7.eps,width=9cm}}
\caption{ $v_{max}$ vs $\tau$ for the Gowdy model with $v_0 = 5$.
The solid curve is from a simulation with 3200 spatial grid points
while the open circles represent one with 20 000 spatial grid points.
The horizontal line is $v_0 = 1$. The difference near the end of the
higher resolution simulation is due to the influence of the 
non-generic point where $Q,_{\theta} = 0$.}
\label{vmax}
\end{center}
\end{figure}

\begin{figure}[bth]
%fig9
\begin{center}
%\setlength{\unitlength}{1cm}
%\makebox[11.7cm]{\psfig{file=gpfig9.eps,width=9cm}}
\makebox[4in]{\psfig{file=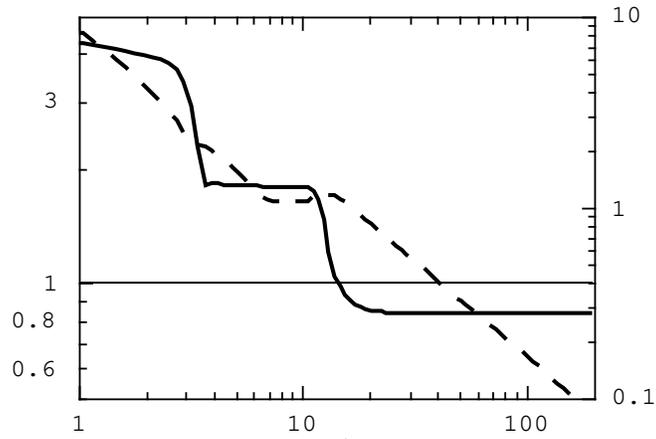,width=3.5in}}
\caption{The same as Figure {\protect \ref{vmax}} with the maximum value
of $P/\tau -v$ over the spatial grid also plotted (dashed line) vs $\tau$.  The
observed behavior is due to the form $P \to v \tau + \ln [(1 + \cos \psi)/2]$ in
the AVTD limit as $\tau \to \infty$.}
\label{pbyt}
\end{center}
\end{figure}

In Fig.~\ref{vmax}, the maximum value over the spatial grid, $v_{max}$, of
the AVTD
parameter $v$ is shown as a function of time $\tau$ for two simulations
with different
spatial resolutions. The steps in $v_{max}$ occur when the location of the
spatial point
with the maximum value of $v$ changes. Note that, for the lower resolution
simulation,
$v_{max}$ eventually falls below unity in support of the GM conjecture \cite{gm1}.
We also see that the higher resolution simulation starts to diverge from the
lower resolution one, signaling the presence of narrow features that are
unresolved by
the coarser resolution.  (The former was not run longer primarily because
the Courant
condition requires small time steps for high resolution studies.) This
suggests (as
confirmed by HS) that the higher spatial resolution simulation will take
longer before
it becomes AVTD everywhere. This is shown explicitly in Fig.~\ref{pipres} where
$\pi_P > 1$ vs $\tau$ is plotted for three different spatial resolutions. Note
that for two of the spikes, $P,_\tau$ is larger for the finest resolution. In
Fig.~\ref{pbyt}, the difference between
$P /
\tau$ and
$v$ is plotted vs.~$\tau$. Since
$P \to v \tau$ as $\tau \to \infty$, we see the influence of the next
(constant in
$\tau$) term $P_0\,+\,\ln[(1\,+\,\cos \psi)/2]$, in the AVTD solution for $P$
(\ref{avtdeq}).

\section{Structure in the Gowdy Model}
\begin{figure}[bth]
%fig1
\begin{center}
%\setlength{\unitlength}{1cm}
%\makebox[11.7cm]{\psfig{file=gpfig1.eps,width=9cm}}
\makebox[4in]{\psfig{file=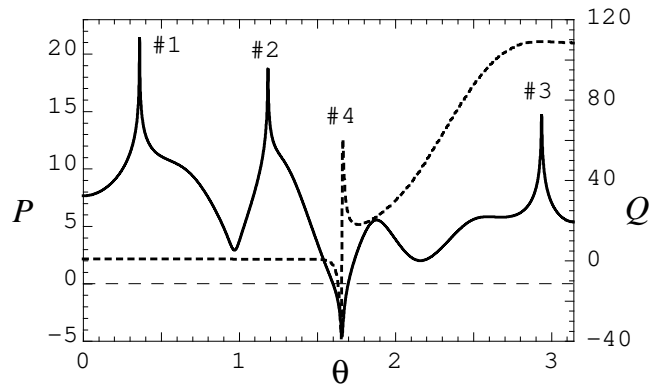,width=3.5in}}
\caption{$P$ (solid line) and $Q$ (dashed line) vs $\theta$ at $\tau = 12.4$ 
for the 
standard initial data set, $P = 0$, $\pi_P = v_0 \, \cos \theta$, $Q=\cos
\theta$, $\pi_Q = 0$, with $v_0 = 5$ for $0 \le \theta \le
\pi$.  The  simulation was run 
with 20000 spatial grid points in the interval $0 \le \theta \le 2 \pi$.  
The numbers on the 
graph refer to the most interesting features.  Peaks \#1, \#2, and \#3 in $P$ 
are essentially the 
same in that they occur at extrema (in $\theta$) of $Q$.  ($Q$ has a maximum 
at the 
locations of peaks \#1 and \#3 and a minimum at \#2.)  Structure \#4 is 
qualitatively different in that there appears to be a discontinuity in $Q$. This  
occurs where $P < 0$ ($P = 0$ is shown as a dashed line) and $\pi_Q \approx 0$.}
\label{gowdyPQ}
\end{center}
\end{figure}

\begin{figure}[bth]
%fig2
\begin{center}
%\setlength{\unitlength}{1cm}
%\makebox[11.7cm]{\psfig{file=gpfig2.eps,width=9cm}}
\makebox[4in]{\psfig{file=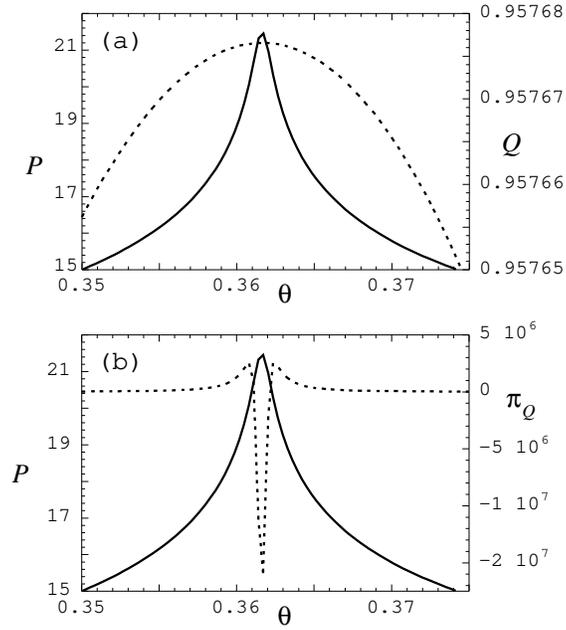,width=3in}}
{\protect \caption{Details of feature \#1 in Figure {\protect \ref{gowdyPQ}}.
(a)
$P$ (solid line) and $Q$ (dashed line) vs $\theta$ near the peak.  Note the
scales in $Q$ and
$\theta$. (b) $P$ (solid line) and $\pi_Q$ (dashed line) vs 
$\theta$ near the peak.}
\label{pqfine}}
\end{center}
\end{figure}

\begin{figure}[bth]
%fig3
\begin{center}
%\setlength{\unitlength}{1cm}
%\makebox[11.7cm]{\psfig{file=gpfig3.eps,width=9cm}}
\makebox[4in]{\psfig{file=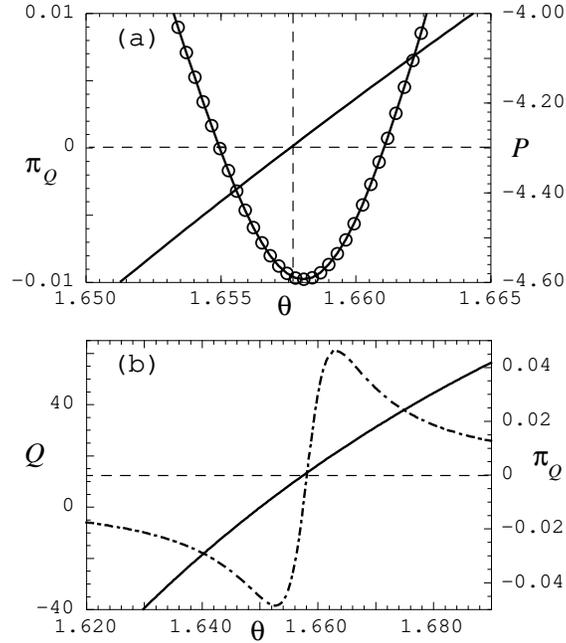,width=3in}}
{\protect \caption{Details of feature \#4 in Figure {\protect \ref{gowdyPQ}}:
(a)
$P$ (circles) and
$\pi_Q$ (solid line) vs $\theta$.  The horizontal dashed line indicates $\pi_Q =
0$ while the vertical dashed line marks the $\theta$ value at which the
$\pi_Q$ goes through zero. Note the offset between the minimum in $P$ and the
zero of $\pi_Q$. (b)
$\pi_Q$ (solid line) and $Q$ (dot-dashed line) vs $\theta$ near the feature. }
\label{qdisc}}
\end{center}
\end{figure}

\begin{figure}[bth]
%fig4
\begin{center}
%\setlength{\unitlength}{1cm}
%\makebox[11.7cm]{\psfig{file=gpfig4.eps,width=9cm}}
\makebox[4in]{\psfig{file=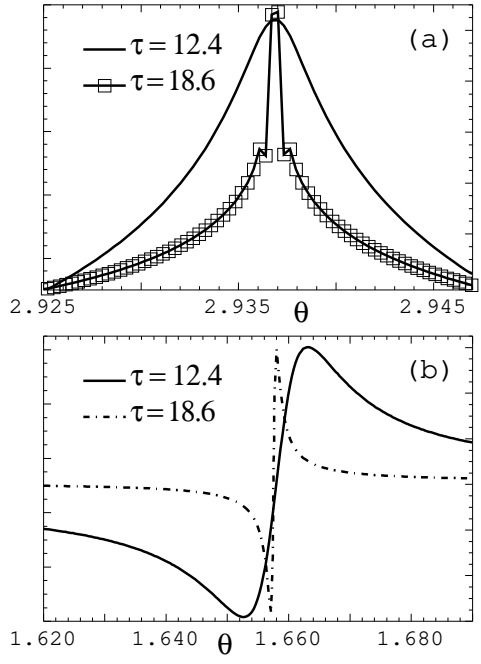,width=2.5in}}
{\protect \caption{Evidence for the narrowing of spiky features with increasing
$\tau$. (a) $P$ as shown in Figure {\protect \ref{pqfine}} is plotted at two
different times. (b)
$Q$ as shown in Figure {\protect \ref{qdisc}} is plotted at two different
times. The amplitudes of $P$ and $Q$ have been rescaled for clarity.}
\label{pqfine18}}
\end{center}
\end{figure}

To understand the Gowdy phenomenology of Section III, we must explore the
structure of the wave equations for $P$ and $Q$. Let us consider
(\ref{gowdywaveQ})
first. The equation for $Q$ may be rewritten in first order form as
\begin{eqnarray}
\label{qequations}
Q,_\tau &=&e^{-2P}\pi _Q\ \ , \nonumber \\
\pi _Q,_\tau &=&e^{-2\tau }\left( {e^{2P}Q,_\theta } \right)\!,_\theta\ .
\end{eqnarray}
This immediately provides the explantion for both the large central peak which
quickly forms in $Q$ and for the apparent discontinuity in $Q$ seen in
Figs.~\ref{gowdyPQ} and \ref{qdisc}. From (\ref{qequations}), $Q$ will grow
rapidly where
$P$ is negative. Since, early in the evolution, $P \approx v_0 \tau \cos
\theta$, one
expects a large peak (as is observed in Figs.~\ref{gowdyPQ} and \ref{pqevol})
where $\cos \theta < 0$. We shall see later that
$P$ is driven to positive values eventually stopping the growth of $Q$. However, we
shall also see that if $\pi_Q \approx 0$, $P$ can remain negative longer. Say $P
\approx - P_0$ for $P_0 > 0$ at a value of $\theta = \theta_1$ where $\pi_Q
\approx \pi_Q^0 (\theta - \theta_1)$. Thus we find that
\begin{equation}
\label{qdiscont}
Q,_\tau \approx \pi_Q^0 (\theta - \theta_1)\,e^{2P_0}
\end{equation}
so that $|Q,_\tau|$ is large and $Q$ increases (decreases) exponentially
depending (say
for $\pi_Q^0 > 0$) on whether $\theta > (<) \theta_1$. Note that we assume
$\pi_Q,_\tau
\approx 0$. However, (\ref{qequations}b) implies this for $P$ large and
negative.

More interesting is the equation for $P$. Consider two
approximate forms of (\ref{gowdywaveP}) depending on which nonlinear term
dominates.
Since the nonlinear terms depend exponentially on $P$, they will dominate the
linear spatial derivative term when the exponential is large.

For example, if $P$ is large and negative,
\begin{equation}
\label{V1pdotdot}
P,_{\tau \tau} - \pi_Q^2 \, e^{-2P} \approx 0
\end{equation}
which has the first integral
\begin{equation}
\label{k1eq}
P,_\tau^2 + \pi_Q^2 \, e^{-2P} = \kappa_1^2.
\end{equation}
This allows one to define an effective potential
\begin{equation}
\label{V1def}
V_1 =  \pi_Q^2 \, e^{-2P}
\end{equation}
which is important when $P,_\tau < 0$ and (especially if) $P < 0$. This allows
immediate explanation of the following phenomena: (a) In the AVTD regime, $P >
0$ and its asymptotic time derivative $v > 0$ since if $P,_\tau < 0$, a bounce
off $V_1$ will occur at which $P,_\tau \to - P,_\tau$ so that $P$ will
eventually reach positive values. (b) However, if $\pi_Q \approx 0$, $V_1
\approx 0$ so that both $P$ and $P,_\tau$ can remain negative for a long time.
Non-generic behavior occurs at the isolated point $\theta = \theta_1$ where
$V_1$ vanishes since $P$ and $P,_\tau$ will never be driven to positive values
there.

On the other hand, if (a) $P$ is large and positive and (b) $P,_\tau > 1$, the
wave equation for $P$ is given by
\begin{equation}
\label{V2pdotdot}
P,_{\tau \tau} + e^{2(P-\tau)}Q,_\theta^2 \approx 0
\end{equation}
which is easy to integrate (for $Z = P - \tau$) as
\begin{equation}
\label{k2eq}
Z,_\tau^2 + Q,_\theta^2 \, e^{2 Z} = \kappa_2^2.
\end{equation}
Thus if $P,_\tau > 1$, the dynamics of $P$ is dominated by an effective
potential
\begin{equation}
\label{V2def}
V_2 = Q,_\theta^2 \, e^{2(P - \tau)}.
\end{equation}
Again, from (\ref{qequations}a), our implicit assumption that $Q,_\tau
\approx 0$ is
consistent. One sees immediately, that if $P,_\tau > 1$ a bounce off $V_2$
will occur
with the result that $Z,_\tau \to - Z,_\tau$ or
\begin{equation}
\label{zbounce}
P,_\tau \to - (P,_\tau - 2).
\end{equation}
Thus the existence of $V_2$ provides a mechanism to drive
$P,_\tau$ and thus asymptotically $v$ to values below unity. In the
actual simulation, the asymptotic regions for the scattering are not
reached so that
(\ref{zbounce}) cannot be used to compute the numerical value of the change
in $P,_\tau$
with $\tau$. Again one sees that non-generic behavior can arise, in this
case, at $\theta
= \theta_2$ where
$Q,_\theta = 0$. At $\theta_2$ the mechanism to drive $v$ below unity is
absent so that
larger values of $v$ are allowed. If $V_2$ is non-vanishing but small, it
will take a
long time for the bounce to occur so that the AVTD limit will take a long time
to arise. This also explains the peaks in $P$ seen in Fig.~\ref{gowdyPQ}. Where $Q,_\theta
\approx 0$, $V_2$ is small so that $P$ can increase without hindrance with a
large positive value of $P,_\tau$.

\begin{figure}[bth]
%fig10
\begin{center}
\makebox[4in]{\psfig{file=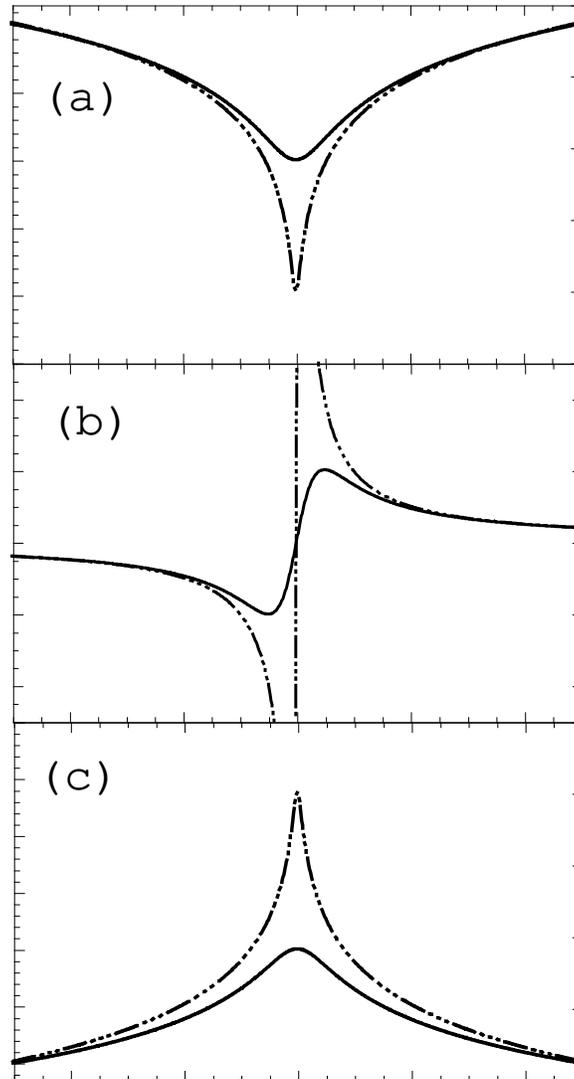,width=3in}}
\caption{Sketches of the analytic approximations to the peaks in $P$ and $Q$.
The larger $\tau$ value is indicated by the broken line. (a) $P$ from (25a).
(b) $Q$ from (25b). (c) $P$ from (27).}
\label{analytic}
\end{center}
\end{figure}

For a more quantitative understanding of the small scale spatial structure,
note that both
equations (18) and (21) have closed form solutions.  Equation (18) holds in
the AVTD regime, where the solution is given in
equation (7).  The exceptional point
$\theta _1$ is the point where $\cos \psi = - 1$.  Near this point we have
$ \cos \psi
\approx - 1 + {{(\psi ')}^2} {{(\theta - {\theta _1})}^2}/2$ where ${\psi
'} \equiv {\psi
_{,\theta }} ( {\theta _1})$.  It then follows that for large $\tau $
\begin{eqnarray}
P \approx {P_0} \; - \; v \tau \; + \; \ln \left [ 1 \, + \, {{(\psi ')}^2}
\, {e^{2 v \tau
}} \, {{(\theta - {\theta _1})}^2}/4 \right ] \nonumber \\
Q \approx {Q_0} \; - \; {{{e^{ - {P_0}}} \, {\psi '} \, (\theta - {\theta
_1})} \over
{2 \, {e^{ - 2 v \tau }} \, + \, {{(\psi ')}^2} {{(\theta - {\theta _1})}^2}/2}}
\end{eqnarray}
Thus there is a spiky feature in both $P$ and $Q$ at $\theta =
{\theta _1}$ as is shown in Fig.~\ref{analytic}.
Furthermore, these features steepen exponentially with time.  Because of
this steepening,
the spatial derivatives of $P$ and $Q$ become large at late times.
However, the AVTD equations are formed by dropping spatial derivatives from
the exact
equations (2) and (3).  Therefore, one might worry that the terms that have been
neglected are
not actually negligible near the peaks.  However, using equation (\ref{PeqGM})
one can show that near
$ \theta = {\theta _1} $ the neglected terms go like $ e^{2 (v-1)\tau}$ and
are therefore
exponentially damped for $v < 1$.

The solution of equation (21) is
\begin{equation}
P = {P_0} \; + \; \tau \; - \; \ln \left [ \cosh w \tau \, - \, \cos \phi
\, \sinh w \tau
\right ] \; \; \; .
\end{equation}
Here, ${P_0}, w \ge 0 $, and $\phi $ are functions of $\theta $, and $ {Q_{,
\theta }} =
{e^{ - {P_0}}} \, w \, \sin \phi $.  Note that $ P \to (1 - w ) \tau $ for
large
$\tau $ where $\cos \phi \ne 1$.  The exceptional point
$\theta _2$ is the point where $\cos \phi = 1$.  Near this point we have $
\cos \phi
\approx  1 - {{(\phi ')}^2} {{(\theta - {\theta _2})}^2}/2$ where ${\phi '}
\equiv {\psi
_{,\theta }} ( {\theta _1})$.  It then follows that for large $\tau $
\begin{equation}
P \approx {P_0} \; + \; (1+w) \tau \; - \; \ln \left [ 1 \, + \, {{(\phi
')}^2} \, {e^{2 w
\tau }} \, {{(\theta - {\theta _2})}^2}/4 \right ].
\end{equation}
Thus the feature in $P$ narrows exponentially with time as can be seen in
Fig.~\ref{analytic}.  The spatial derivatives of $P$
become very large.  However, the neglected terms in the equations go like
$e^{2(w-1)\tau }$.
Since $ w - 1 < 0 $ for $ {P_{, \tau }} > 0 $ asymptotically, it follows
that these terms
become negligible after the last bounce.

\begin{figure}[bth]
%fig8
\begin{center}
%\setlength{\unitlength}{1cm}
%\makebox[11.7cm]{\psfig{file=gpfig8.eps,width=9cm}}
\makebox[4in]{\psfig{file=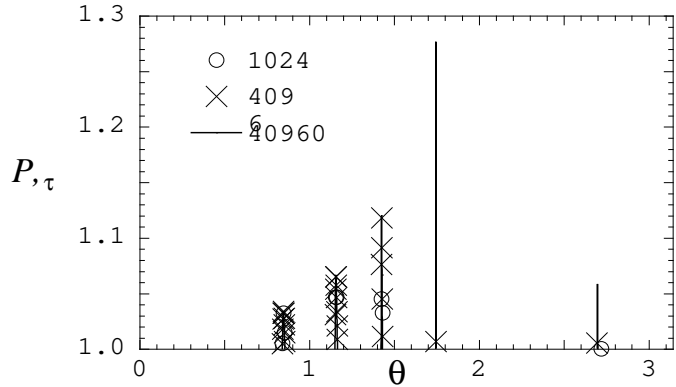,width=3.5in}}
\caption{$P,_\tau > 1$ vs $\theta$ at $\tau = 24.79$ with standard initial
data with $v_0 = 10$ for three spatial resolutions. For the coarser
resolutions, all simulation spatial points are shown.}
\label{pipres}
\end{center}
\end{figure}

\begin{figure}[bth]
%fig11
\begin{center}
\makebox[4in]{\psfig{file=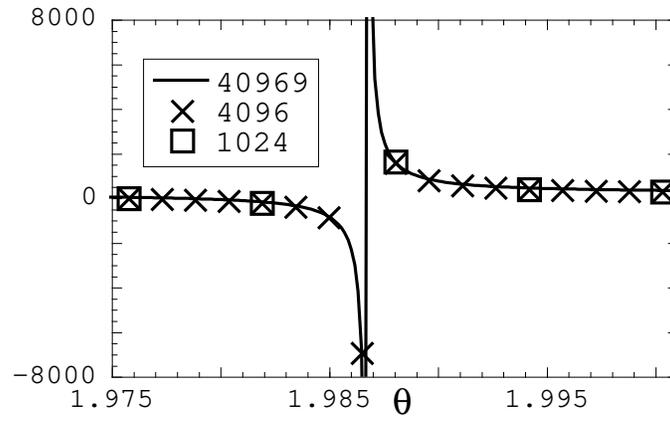,width=3.5in}}
\caption{$Q$ vs $\theta$ for $\tau = 24.79$ with $v_0 = 10$ at the site of the
apparent discontinuity for three spatial resolutions. At this value of $\tau$,
the feature is not resolved even at the finest resolution used.}
\label{qspikeres}
\end{center}
\end{figure}

\begin{figure}[bth]
%fig12
\begin{center}
\makebox[4in]{\psfig{file=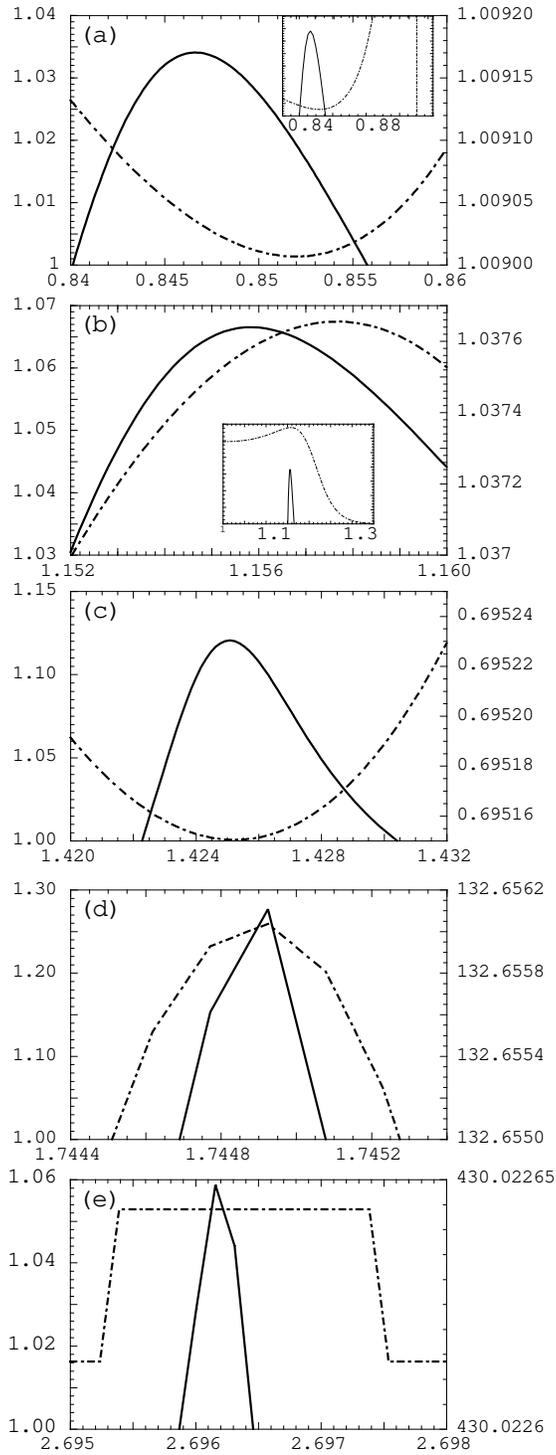,width=3in}}
\caption{Details of $P$ (solid line, left axis scale) vs $Q$ (dot-dashed line,
right axis scale) for the spikes in Figure {\protect \ref{pipres}}. (a) Note
that the spike is completely resolved and that there is an offset of the
extrema. The inset shows that this peak is part of a larger feature in $Q$. (b)
Again the offset can be related to a larger feature in $Q$. (c) The extrema are
much more closely aligned. (d) The features are not completely resolved. The
extrema are aligned to within the resolution. (e) Again the features are not
well resolved. The change in the value of $Q$ is the smallest that can be seen
in single precision. The simulations were done in double precision but the data
was reported in single precision.}
\label{pipvsq}
\end{center}
\end{figure}

\begin{figure}[bth]
%fig13
\begin{center}
\makebox[4in]{\psfig{file=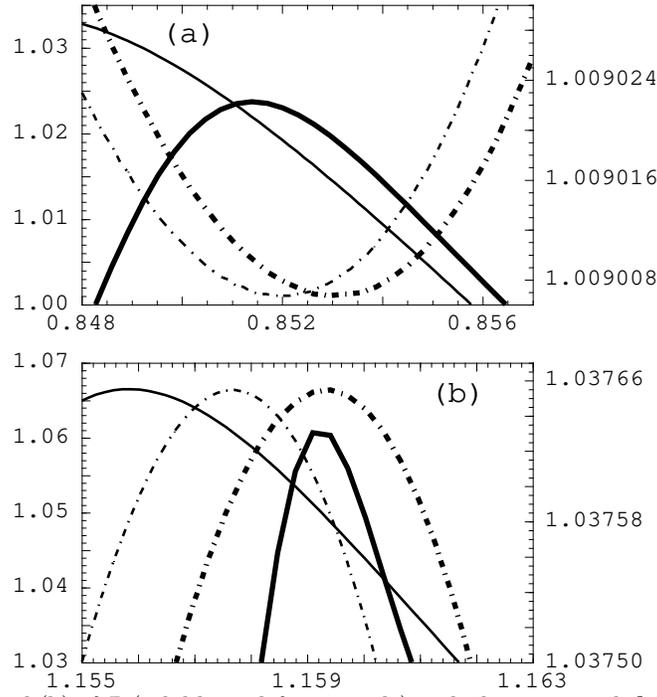,width=3.5in}}
\caption{Spiky features (a) and (b) of $P$ (solid lines, left axis scale) with
the associated
$Q$ (dot-dashed lines, right axis scale) in Figure {\protect \ref{pipvsq}} at
$\tau = 49.5$. Part of the previous figure (at $\tau = 24.8$) is shown with
finer lines. Note the improvement in alignment and the narrowing of the spike
in $P$.}
\label{pipreslater}
\end{center}
\end{figure}

The differences in Figs.~\ref{vmax} and \ref{pipres} seen by increasing the
spatial
resolution are also understood in terms of our analysis of structure growth. At
higher spatial resolution it is more likely that a
spatial grid point of the numerical simulation will be very close to a
non-generic point
where $Q,_\theta = 0$. Thus the non-generic behavior will become more
apparent and the
time to reach the AVTD limit characterized by $v < 1$ everywhere will be longer.
Clearly, the observed resolution dependence may be interpreted to be
evidence for the
existence of non-generic points. The apparent discontinuity in $Q$ seen in
Figs.~\ref{gowdyPQ} and \ref{qdisc} is evidence for the non-generic point in
$\theta$
where $\pi_Q = 0$. Its effect may be seen as a resolution dependence of the
shape of this
feature. It becomes steeper and narrower with increasing spatial resolution for
sufficiently large $\tau$ (see Fig.~\ref{qspikeres}) or for increasing $\tau$
with fixed spatial resolution (see Fig.~\ref{pqfine18}b). As in
Fig.~\ref{pqfine}, we note that the spikes in Fig.~\ref{pipres} where
$v > 1$ are associated with extrema in $Q$. This is shown in Fig.~\ref{pipvsq}.
Note that the alignment is not perfect between the peak in $P,_\tau$ and the
extremum in $Q$ where the spike is well resolved (see also
Fig.~\ref{qdisc}a). This is understood as evidence that the AVTD regime has not
yet been reached. In particular, the site of the extremum may be distorted by a
nearby larger feature in $Q$. Running the simulation twice as long
(Fig.~\ref{pipreslater}) shows much closer alignment and is evidence that the
predicted non-generic behavior will eventually dominate the spikes.

\begin{figure}[bth]
%fig14
\begin{center}
\makebox[4in]{\psfig{file=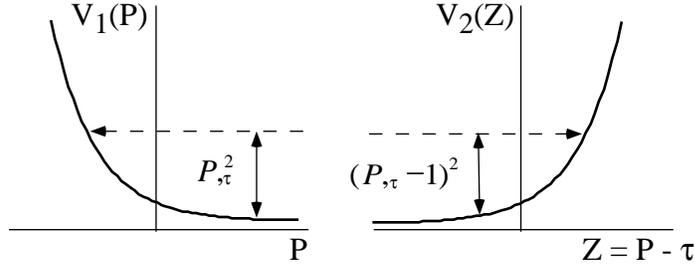,width=4in}}
\caption{$V_1$ vs $P$ and $V_2$ vs $P - \tau$ shown as solid lines. The
horizontal dashed arrows denote the constants $\kappa_1$ and $\kappa_2$.}
\label{potentials}
\end{center}
\end{figure}

\begin{figure}[bth]
%fig15
\begin{center}
\makebox[4in]{\psfig{file=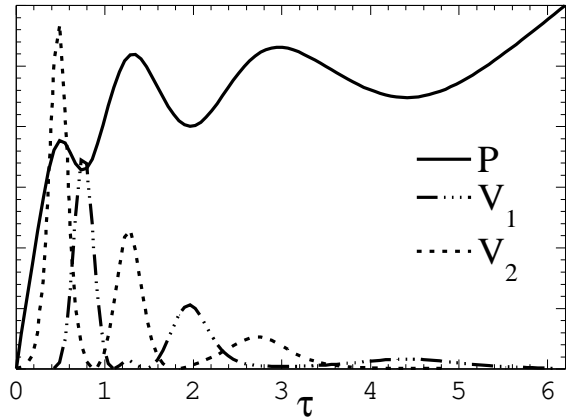,width=3in}}
\caption{$P$ (solid curve), $V_1$ (dash-dotted curve), and $V_2$ (dashed curve) vs
$\tau$ at a fixed value of $\theta$.}
\label{pv1v2}
\end{center}
\end{figure}

\begin{figure}[bth]
%fig16
\begin{center}
\makebox[4in]{\psfig{file=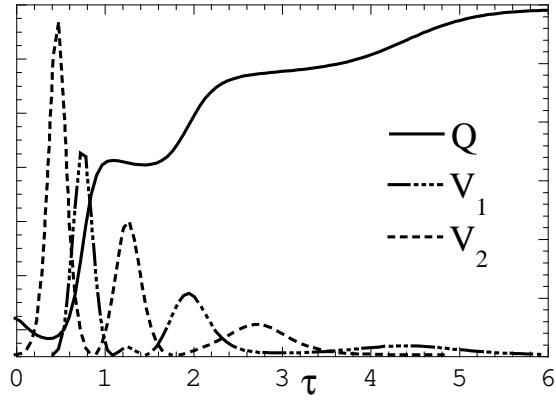,width=3in}}
\caption{$Q$ (solid curve), $V_1$ (dash-dotted curve), and $V_2$ (dashed curve)
vs $\tau$ at the same value of $\theta$ as in Figure {\protect \ref{pv1v2}}.}
\label{qv1}
\end{center}
\end{figure}

\begin{figure}[bth]
%fig17
\begin{center}
\makebox[4in]{\psfig{file=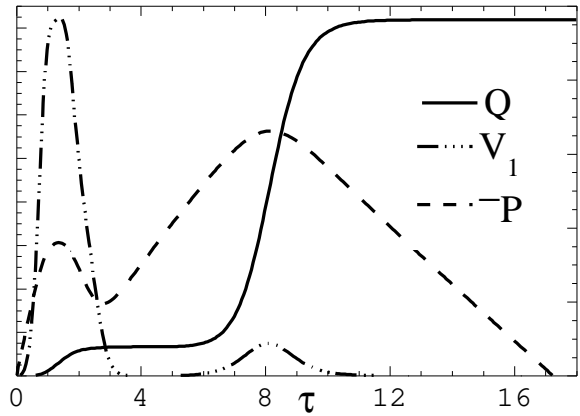,width=3in}}
\caption{$Q$ (solid curve), $V_1$ (dash-dotted curve), and $-P$ (dashed curve)
vs
$\tau$ at the site of feature \#4 in Figure {\protect \ref{gowdyPQ}}.}
\label{qminusp}
\end{center}
\end{figure}

The effective potentials $V_1$ and $V_2$ are displayed in
Fig.~\ref{potentials}. We see
that the approach to the AVTD regime is characterized by repeated alternate
interactions of $P$ with the two potentials. This is displayed explicitly for a
typical value of $\theta$ in Fig.~\ref{pv1v2} where $P$, $V_1$, and $V_2$
are shown as
functions of $\tau$. After each interaction with $V_2$, the quantity
$P,_\tau - 1$ changes
sign. If this yields $0 \le P,_\tau < 1$, no further bounces occur. If
$P,_\tau < 0$, there
will be an interaction with $V_1$. Eventually, $|P,_\tau|$ falls below
unity so that
$V_2$ permanently disappears. $P$ then continues to increase for all
subsequent $\tau$
with constant slope $0\le v < 1$ as is characteristic of the AVTD limit.
(Note that
$V_1$ will also decrease exponentially as $P$ increases even though it is part
of the AVTD solution.) Examination of (\ref{qequations}) shows that, when
$|P,_\tau|$ falls below
unity and $\tau \to \infty$, $\pi_Q,_\tau \to 0$ as required by the AVTD
limit. Once $\pi_Q$ becomes constant, as $P$ increases, $Q,_\tau \to 0$, again
as required. The development of $Q$ with time is shown in
Fig.~\ref{qv1} for the same
value of $\theta$ as in the previous figure. Here we see clearly how $Q$ grows
fastest when $V_1$ dominates. This becomes even more dramatic in
Fig.~\ref{qminusp}
which is the same graph for $\theta \approx \theta_1$, the site of the
apparent discontinuity in $Q$.

It is important to emphasize that the magnitude and time evolution of $V_1$
and $V_2$ as well as the initial values of the wave amplitude time
derivatives are
different at different values of
$\theta$. This explains the development of complicated small scale spatial
structure.
From (\ref{k1eq}) or (\ref{k2eq}), one computes the time to the first
bounce off $V_1$ or
$V_2$ as
\begin{equation}
\label{bouncetime}
\Delta \tau =\int_0^{- \ln ({{\sqrt {V_J^0}} \mathord{\left/ {\vphantom {{\sqrt
{V_J^0}} {\kappa _J)}}} \right. \kern-\nulldelimiterspace} {\kappa _J)}}}
{{{d{\cal
X}_J} \over {\kappa _J\sqrt {1-{{V_J} \mathord{\left/ {\vphantom {{V_J} {\kappa
_J^2}}} \right.
\kern-\nulldelimiterspace} {\kappa _J^2}}}}}}
\end{equation}
where $J=1$ or $2$ and ${V^0 _J}={V_J}$ at $\tau = 0$.  Here ${\chi _1} = - P$
and
${\chi _2} = Z$.
Evaluation of (\ref{k1eq}) and (\ref{k2eq}) at $\tau
= 0$ yields an approximate inverse scaling with $v_0$ seen in
Fig.~\ref{scaling}. The
central peak appears first and is easy to understand. Since, initially,
$P,_\tau = v_0
\cos \theta$, the highest negative velocity occurs at
$\theta = \pi$ causing the bounce off $V_1$ to occur there first. $P$ then
begins to
increase at $\theta = \pi$ both earlier and faster than at neighboring values of
$\theta$. More complicated is the next pair of peaks which occur near $\theta =
\pi/4,\,\, 7 \pi / 4$. Initially, the profile of $P$ vs $\theta$ steepens.
However,
bounces off $V_2$ will begin to occur. This causes the tangent to $P$ vs
$\theta$ to
evolve from negative (positive) to positive (negative) slope for $0 \le \theta
\le \pi/2$ ($3 \pi / 2 \le \theta \le 2 \pi$) eventually creating a pair of
peaks. The
higher the original value of $v_0$, the greater will be the number of
bounces before
$V_2$ disappears for good. Ever finer small scale structures are created at
$\theta$
values where many bounces occur since $P$ will increase for some spatial
regions and
decrease for others in a manner that changes with time. As $v$ falls below
unity, at a
given $\theta$, the bounces, and thus structure formation, will cease there but
structure formation will continue in the ever smaller spatial regions where $v$
continues to exceed unity.

\section{Discussion and Conclusions}
We see that both potentials $V_1$ and $V_2$ drive the cosmology toward the
AVTD regime
characterized by (\ref{avtdeq}) as $\tau \to \infty$ with $0 \le P,_\tau
\to v < 1$. This
is because $V_1$ appears and influences the dynamics only when and where
$P,_\tau < 0$
while $V_2$ does so only for $P,_\tau > 1$. In the AVTD regime, with $0 \le
v < 1$, both
corresponding terms in the wave equation for $P$ are exponentially small.
We note that
the wave equations (\ref{gowdywaveP}) and (\ref{gowdywaveQ}) provide no
other mechanism
to drive
$P,_\tau$ into the required range if it is not there initially. Recall that the
polarized Gowdy model can have any value for $P,_\tau$. Thus we argue that
at isolated
points (a set of measure zero in $\theta$) where either $\pi_Q$ or
$Q,_\theta$ is
precisely zero,
$P,_\tau$ could remain outside the allowed range. In the former case (say
at $\theta =
\theta_1$), the consistency equations (\ref{PeqGM}) are satisfied so that one
expects the solution at
$\theta_1$ to be like an AVTD polarized solution. However, in the latter
case (say at
$\theta = \theta_2$), the consistency conditions require $Q,_{\theta
\theta} = 0$ as
well as $Q,_\theta = 0$---something one does not expect generically. If $Q,_{\theta
\theta} = 0$ at $\theta_2$, then the solution is of the AVTD large $v$ type
there. If
$Q,_{\theta \theta } \ne 0$, the behavior cannot be analyzed either
numerically or by
these analytic methods. We may therefore conclude that the simulations
provide strong
support for the conjecture that the Gowdy models have a singularity which
is AVTD
except, perhaps, at a set of measure zero.

The existence of this set of measure zero of ``non-generic'' spatial points
can be
inferred from the simulations since near them the potentials are very flat.
Thus, the
closer one is to e.g. $\theta_1$ or $\theta_2$ defined as above, the longer
it will
take to drive $P,_\tau$ to the range $[0,1)$ there. This manifests itself as a
resolution dependence of the simulations of a very specific type. At any
value of $\tau$
(say $\bar \tau$), there is a spatial resolution sufficient to resolve all spiky
features. (This was demonstrated in \cite{bkbvm}.) But the ``center'' of
each spiky
feature is the site of a non-generic point. As one moves away from this
point in either direction in $\theta$, the coefficient of the potential
will be larger.
Thus the interaction with the potential will occur first at regions away
from the center
of the feature and move closer to the center as $\tau$ increases.
Since the
effect of the interaction is to change the sign of either $P,_\tau$ or
$P,_\tau -1$,
the interactions will tend to eliminate the outer part of the spiky
feature. Thus the
spiky feature gets narrower in time until for some $\tau > \bar
\tau$, the original resolution is no longer adequate. Although, in a peak near
$Q,_\theta = 0$, $P,_\tau$ should be $> 1$, inadequate resolution may act
as an effective
averaging to give a measured smaller value of
$P,_\tau$. Thus the peak in $v$ appears to decrease in amplitude only
because, at most
$\theta$ values, the interaction with $V_2$ has already occurred. We note
that this
effective averaging will give $0 \le v < 1$ everywhere for some $\tau_F$
for any fixed
spatial resolution. As the spatial resolution is increased, $\tau_F$ will also
increase since accurate resolution of spiky features can be maintained
farther into
the simulation and the effective averaging will take longer to come into
play. Thus
the apparent resolution dependence is consistent with the behavior of the
potentials
$V_1$ and $V_2$ expected near the non-generic points.

\begin{figure}[bth]
%fig18
\begin{center}
\makebox[4in]{\psfig{file=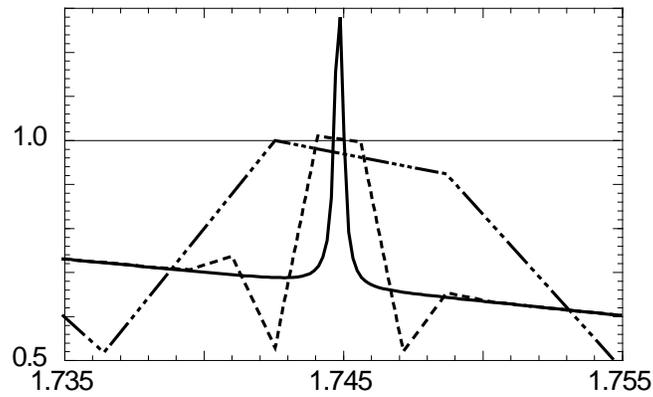,width=3.5in}}
\caption{The same spiky feature in $P,_\tau$ vs $\theta$ is shown at the same
$\tau$ for  resolutions of 1024 (dash-dotted line), 4096 (dashed line), and
40960 (solid line) spatial grid points. The horizontal line marks $v = 1$.}
\label{ptau2res}
\end{center}
\end{figure}

\begin{figure}[bth]
%fig19
\begin{center}
\makebox[4in]{\psfig{file=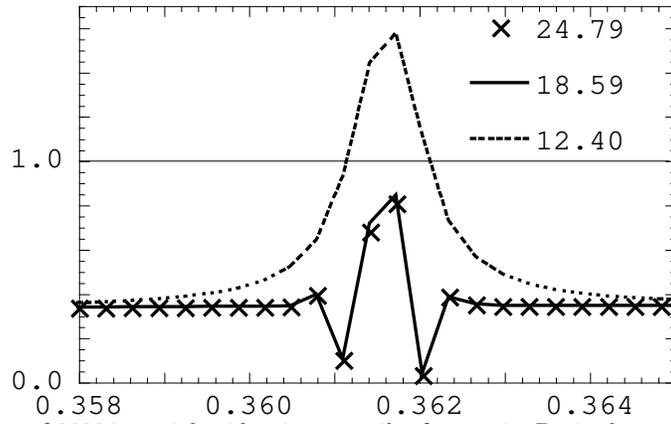,width=3.5in}}
\caption{For a fixed resolution of 20224 spatial grid points, a spiky feature
in $P,_\tau$ is shown vs $\theta$ for three different values of $\tau$.}
\label{threetaus}
\end{center}
\end{figure}

\begin{figure}[bth]
%fig20
\begin{center}
\makebox[4in]{\psfig{file=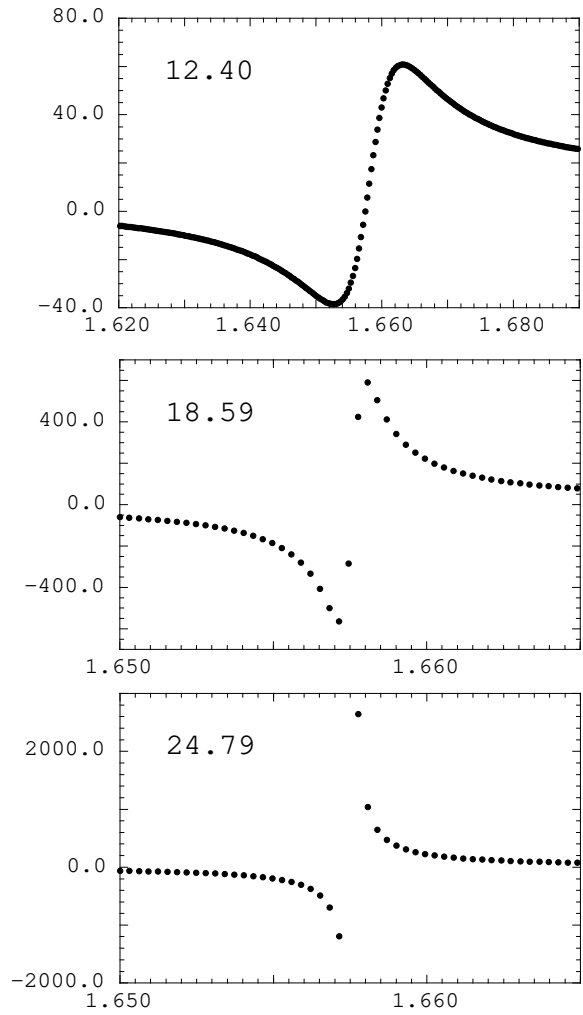,width=3in}}
\caption{The apparent discontinuity in $Q$ vs $\theta$ at three values of
$\tau$ with 20224 spatial grid points.}
\label{qnarrow}
\end{center}
\end{figure}

\begin{figure}[bth]
%fig21
\begin{center}
\makebox[4in]{\psfig{file=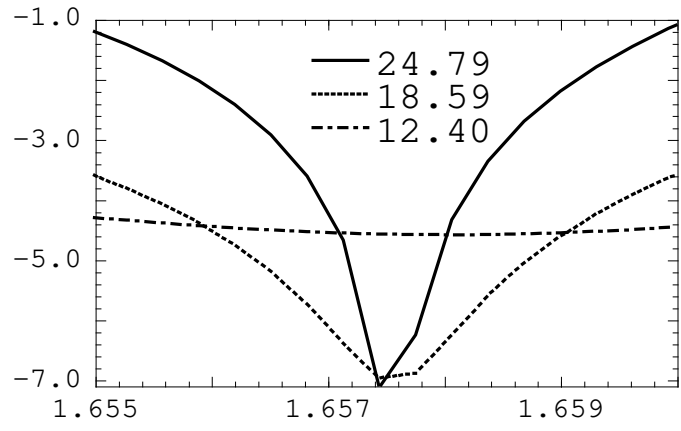,width=3.5in}}
\caption{Part of the region where $P < 0$ vs $\theta$ at three values of $\tau$
for 20224 spatial grid points.}
\label{pneg}
\end{center}
\end{figure}

Several examples are shown in Figs.~\ref{ptau2res}--\ref{pneg}. In
Fig.~\ref{ptau2res}, a
spiky feature in
$P,_\tau$ is shown for three different spatial resolutions at $\tau =
24.79$. While
the coarsest resolution shows $v < 1$, it also clearly fails to resolve the
spiky feature.
The highest resolution both resolves the spiky feature and shows that it has
a core where
$v > 1$ as expected. A similar spiky feature in $P,_\tau$ is shown in
Fig.~\ref{threetaus} for three different times for the same spatial
resolution. One
expects that the clearly resolved spike observed early in the simulation
will narrow as
$\tau$ increases remaining always
$> 1$ at the center of the peak. This is not observed (rather one sees the
constant $P,_\tau$ behavior expected in the AVTD limit). However, one again sees
clearly that the spatial resolution which was adequate at $\tau = 12.4$ has
become inadequate at
later times and that the properties of the spike are not correctly represented.
Fig.~\ref{qnarrow} shows that the apparent discontinuity in $Q$ becomes
narrower and
steeper with time and eventually is not adequately resolved at a given spatial
resolution. One expects $P$ to be increasingly negative at the center of
this feature
since $P,_\tau < 0$ is maintained at that non-generic point. At fixed spatial
resolution, we see in Fig.~\ref{pneg} that the negative region of $P$ both
narrows and
deepens as $\tau$ increases from $12.4$ to $18.6$ as expected. However, at
$\tau =
24.8$, the expected behavior is not seen since the depth of the negative
region fails to
increase. However, we also see that the spatial resolution has become
inadequate.
Presumably, adequate spatial resolution would show the narrowness and depth
of the
negative region suggested by the curvature of $P$ vs $\theta$ in the region just outside
the core.

In conclusion, the behavior of generic Gowdy cosmologies on $T^3 \times R$
can be
completely understood in terms of the nonlinear interactions between the two
polarizations of the gravitational waves. These act as effective potentials
which drive
the system to the prediced AVTD behavior and then cease to play a role. At
the set of
measure zero where the potentials are absent, $P,_\tau$ may lie outside its
allowed
AVTD range of $[0,1)$ for all $\tau$ if it does so at any time since the
mechanism to
correct its value is absent. The existence of this set of measure zero is
inferred from
the details of the shape and time dependence of the spiky features and from
the inability
to resolve them with a given spatial resolution beyond a certain time.

\section*{Acknowledgements}
The authors would like to thank Vincent Moncrief and Boro Grubi\u{s}i\'{c}
for useful
discussions. BKB would like to thank  the Institute for Geophysics and
Planetary Physics
at Lawrence Livermore National Laboratory and the Albert Einstein Institute
at Potsdam
for hospitality. This work was supported in part by National Science Foundation
Grants PHY9507313 and PHY9722039 to Oakland University. Computations were
performed at the National Center for Supercomputing Applications (University of
Illinois).

\pagebreak

\end{document}